\documentclass[aps,notitlepage,twocolumn,superscriptaddress,prl,10pt]{revtex4-2}

\usepackage[utf8]{inputenc}
\usepackage{dsfont}
\usepackage{latexsym,revsymb}
\usepackage{todonotes}
\usepackage{times}

\usepackage[matrix,frame,arrow]{xy}
\usepackage{amsmath}

\newcommand{\ket}[1]{\left\vert{#1}\right\rangle}
\newcommand{\qw}[1][-1]{\ar @{-} [0,#1]}

\newcommand{\gate}[1]{*{\xy *+<.6em>{#1};p\save+LU;+RU **\dir{-}\restore\save+RU;+RD **\dir{-}\restore\save+RD;+LD **\dir{-}\restore\POS+LD;+LU **\dir{-}\endxy} \qw}

\newcommand{\measureD}[1]{*{\xy*+=+<.5em>{\vphantom{\rule{0em}{.1em}#1}}*\cir{r_l};p\save*!R{#1} \restore\save+UC;+UC-<.5em,0em>*!R{\hphantom{#1}}+L **\dir{-} \restore\save+DC;+DC-<.5em,0em>*!R{\hphantom{#1}}+L **\dir{-} \restore\POS+UC-<.5em,0em>*!R{\hphantom{#1}}+L;+DC-<.5em,0em>*!R{\hphantom{#1}}+L **\dir{-} \endxy} \qw}

\newcommand{\multigate}[2]{*+<1em,.9em>{\hphantom{#2}} \qw \POS[0,0].[#1,0];p !C *{#2},p \save+LU;+RU **\dir{-}\restore\save+RU;+RD **\dir{-}\restore\save+RD;+LD **\dir{-}\restore\save+LD;+LU **\dir{-}\restore}

\newcommand{\ghost}[1]{*+<1em,.9em>{\hphantom{#1}} \qw}

\newcommand{\ustick}[1]{*!D!<0em,-.2em>=<0em>{{\scriptstyle #1}}}

\newcommand{\Qcircuit}[1][0em]{\xymatrix @*=<#1>} 

\newcommand{\prepareC}[1]{*{\xy*+=+<.5em>{\vphantom{#1\rule{0em}{.1em}}}*\cir{l^r};p\save*!L{#1} \restore\save+UC;+UC+<.5em,0em>*!L{\hphantom{#1}}+R **\dir{-} \restore\save+DC;+DC+<.5em,0em>*!L{\hphantom{#1}}+R **\dir{-} \restore\POS+UC+<.5em,0em>*!L{\hphantom{#1}}+R;+DC+<.5em,0em>*!L{\hphantom{#1}}+R **\dir{-} \endxy}}

\usepackage{mathtools}
\usepackage{amsthm}
\usepackage{enumitem}
\usepackage{amsmath,amssymb,mathrsfs}
\usepackage{hyperref}
\usepackage{graphicx}
\usepackage{color}

\hypersetup{breaklinks=true}

\newcommand{\transf}[1]{\ensuremath{\mathcal{#1}}}
\newcommand{\tA}{\transf A}

\newcommand{\tC}{\transf C}
\newcommand{\tD}{\transf D}

\newcommand{\tM}{\transf M}

\newcommand{\tS}{\transf S}
\newcommand{\tT}{\transf T}
\newcommand{\tU}{\transf U}
\newcommand{\tV}{\transf V}
\newcommand{\tW}{\transf W}
\newcommand{\tX}{\transf X}

\newcommand{\tI}{\transf I}

\newcommand{\sys}[1]{\ensuremath{\mathrm{#1}}}
\newcommand{\rA}{\sys A}
\newcommand{\rB}{\sys B}
\newcommand{\rC}{\sys C}

\newcommand{\rE}{\sys E}
\newcommand{\rF}{\sys F}

\newcommand{\st}{\mathsf{St}}

\DeclarePairedDelimiter{\norma}{\lVert}{\rVert}

\DeclarePairedDelimiter{\normad}{\lVert}{\rVert_{\diamond}}

\newcommand{\Tr}{\mathrm{Tr}}

\newcommand{\proj}[1]{|{#1}\rangle\!\langle{#1}|}
\def\>{\rangle}

\def\<{\langle}

\def\Tr{\operatorname{Tr}}

\newcommand{\QCs}[1]{\mathfrak{C}(#1)}
\newcommand{\QC}[2]{\mathfrak{C}(#1,#2)}

\newcommand{\QUs}[1]{\mathfrak{U}(#1)}

\begin{document}

\title{Causal influence versus signalling for interacting quantum
  channels}

\author{Kathleen \surname{Barsse}}

\email{kathleen.barsse@ens-paris-saclay.fr}

\affiliation{Universit\'e Paris-Saclay, ENS Paris-Saclay, 91190, Gif-sur-Yvette, France}

\affiliation{QUIT group, Physics Dept., Pavia University, and INFN Sezione di Pavia, via Bassi 6, 27100 Pavia, Italy}

\author{Paolo \surname{Perinotti}}

\email{paolo.perinotti@unipv.it}

\affiliation{QUIT group, Physics Dept., Pavia University, and INFN Sezione di Pavia, via Bassi 6, 27100 Pavia, Italy}

\author{Alessandro \surname{Tosini}}

\email{alessandro.tosini@unipv.it}

\affiliation{QUIT group, Physics Dept., Pavia University, and INFN Sezione di Pavia, via Bassi 6, 27100 Pavia, Italy}

\author{Leonardo \surname{Vaglini}}

\email{leonardo.vaglini01@universitadipavia.it}

\affiliation{QUIT group, Physics Dept., Pavia University, and INFN Sezione di Pavia, via Bassi 6, 27100 Pavia, Italy}

\begin{abstract}
  A causal relation between quantum agents, say Alice and Bob, is
  necessarily mediated by an interaction. Modelling the last one as a
  reversible quantum channel, an intervention of Alice can have causal
  influence on Bob's system, modifying correlations between Alice and
  Bob's systems. Causal influence between quantum systems necessarily
  allows for signalling. Here we prove a mismatch between causal
  influence and signalling via direct computation of the two
  quantities for the $\mathrm{Cnot}$ gate. Finally we show a
  continuity theorem for causal effects of unitary channels: a
  channel has small causal influence iff it allows for small signalling.
\end{abstract}

\maketitle

Establishing causal relationships is a primary issue in 
science~\cite{pearl_2009,PhysRevA.88.052130} as well as to use causal 
relations to infer information on the underlying processes~\cite{PhysicsPhysiqueFizika.1.195,PhysRevLett.23.880,Pawowski:2009aa,PhysRevX.7.031021,Renou:2021aa}. Prompted by the study of
all facets of quantum nonlocality, the causal structure of networks
of quantum systems has been largely explored in the light of
information theory~\cite{Popescu1994, Beckman:2001aa,Bennett-2003,Eggeling:2002aa,PhysRevA.74.012305,Schumacher:2005aa, PhysRevA.72.062323,GutoskiWatrous-testers,PhysRevA.80.022339, PhysRevLett.101.060401,PhysRevA.94.032131},
paving the way towards recent developments in the direction of
quantum indefinite causal
order~\cite{Hardy2009,PhysRevA.88.022318,Oreshkov2012aa,Brukner-qc,Oreshkov_2016,7867830,Perinotti2017,bisio2019theoretical,10.1145/3581760,arrighi2022quantum}. In the latter context,
also the order of processes is taken as a quantum degree of freedom, thus unlocking new resources~\cite{PhysRevLett.113.250402,Milz2022resourcetheoryof,PhysRevLett.117.100502,Chiribella_2021}.

The question at the core of quantum causal
models~\cite{Barrett-qcm,Perinotti2021causalinfluencein}
is whether a dynamics, which in an informational setting corresponds to
a gate with composite input and output systems, can or cannot induce
cause-effect relations between the involved parties. In this respect,
much of the attention so far was given to \emph{communication}, studying the
structure of quantum gates in relation to their capacity to exchange information
between timelike separated 
parties~\cite{Beckman:2001aa,Bennett-2003,Eggeling:2002aa,PhysRevA.74.012305}. While 
the role of communication captures only one instance of causal
relations~\cite{Perinotti2021causalinfluencein},
it makes it clear that non trivial causal effects between far apart
systems must be mediated by an
\emph{interaction}~\cite{Beckman:2001aa}. In the absence of an
interaction, one can indeed trivially assume that there is no \emph{causal
influence} between two systems. On the other hand, the presence of an
interaction mediates a causal influence, that manifests itself in the
creation of \emph{correlations} between the interacting systems.

A largely unexplored side of quantum information processing is the
scaling of causal effects versus the strength of the ``coupling''
between the systems involved. In the same line of thought, the
relation between the strength of correlations and the amount of
signalling is clearly a question of interest, that is largely unexplored.

In the present Letter we address the last question by defining
quantifiers of signalling and causal influence and studying relations
between them. It turns out that, just as it holds {\em no signalling
  if and only if no causal
  influence}~\cite{Perinotti2021causalinfluencein}, one has the
stronger condition {\em little signalling if and only if little causal
  influence}. The continuity bounds for casual influence and
signalling leave room for differences in the two quantities. Indeed,
we show that, for the quantum $\mathrm{Cnot}$, signalling is strictly
smaller than causal influence, thus indicating that the ``extra''
causal effect beyond signalling has to be sought in the leverage that
it enables on correlations.

For simplicity, we will restrict to finite dimensional systems, and
use the same capital Roman letter $\rA$ to denote a quantum system and
the corresponding Hilbert space. In quantum theory an interaction
between two systems, say $\rA$ (controlled by Alice) and $\rB$
(controlled by Bob) is represented by a bipartite channel $\tC$ (i.e.,
a completely positive trace preserving map) sending quantum states of
the Hilbert space $\rA\otimes\rB$ to quantum states of the Hilbert
space $\rA'\otimes\rB'$, with $\otimes$ denoting the usual Hilbert
spaces tensor product. We will write $\QC{\rC}{\rC'}$ for the set of
channels from $\rC$ to $\rC'$ (shortened to $\QCs{\rC}$ when
$\rC'=\rC$), and $\QUs{\rC}$ for that of {\em unitary} channels on system
$\rC$.  We will adopt the following graphical representation for
bipartite quantum channels
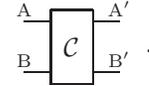
\begin{equation*}
	\begin{aligned}
	\Qcircuit @C=1em @R=1em
	{&\ustick{\rA}&\multigate{1}{\tC}&\ustick{\rA'}\qw\\
		&\ustick{\rB}&\ghost{\tC}&\ustick{\rB'}\qw}
\end{aligned}\quad .
\end{equation*}
The preparation of a state $\rho$ and the measurement
of a POVM $\{M_x\}$ on system $\rA$ will be
graphically represented as
$\Qcircuit @C=1em @R=1em { \prepareC{\rho} &\ustick{\rA}\qw }\:\:$\, and
\, $\:\:\Qcircuit @C=1em @R=1em { \ustick{\rA} & \measureD{M_x}}$ ,
respectively. We will denote the set of states of system
$\rA$ by the symbol $\st(\rA)$.

A bipartite channel $\tC\in\QC{\rA\rB}{\rA'\rB'}$ models an
interaction between the quantum systems of the two users,
Alice and Bob, and we will analyse it in terms of causal relations that it produces 
between Alice's input and Bob's output. As noticed by
several authors~\cite{barrett2019quantum,Perinotti2021causalinfluencein}, the study of causal relations generated by non
reversible channels may be ambiguous. Indeed, any channel $\tC$ can be
realized in a non unique way as a reversible channel $\tU$ by
discarding an appropriate environment system. It happens that the occurrence of
causal relations between agents actually depends of the specific initial state of the environment involved in a reversible dilation. Accordingly, causal relations are unambiguously identified once the description is expanded such that all relevant
systems are included, thus dealing with an ``isolated system". For this
reason we focus on the evolution of isolated quantum systems, thus
exploring the causal relations mediated by unitary channels.

One extreme case is that where systems $\rA$ and $\rB$ are separately isolated, thus
non interacting. Clearly, in this case the evolution channel cannot produce any causal 
relation between Alice and Bob. On the other hand, if one e.g.~swaps systems $\rA$ and 
$\rB$, the result is that the swap channel mediates as much causal influence as one can 
possibly expect. Now, still on the same line of thought, one can expect that a ``little'' 
interaction induces ``little'' causal effects. 
However, in order to prove this intuition one needs to introduce first suitable 
quantifiers for interaction and causal influence. 


We then start by introducing two 
functions on the set of quantum unitary channels, denoted by
$S(\tU)$ and $C(\tU)$, that quantify the
amount of signalling and that of causal influence from Alice to Bob 
for the channel $\tU$, respectively.

Signalling, that is
communication from Alice to Bob (or vice-versa), is based on the dependence of the local
output system $\rB'$ of Bob's on the choice of the local input
system $\rA$ of Alice's: in general, Alice can influence the outcome probabilities for 
Bob's local measurements on $\rB'$, by varying her choice of intervention on 
system $\rA$. If Bob's output at $\rB'$ does not depend on the
state of Alice's input at $\rA$, then we say that $\tU$ is
\emph{no-signalling} from Alice to Bob. One can straightforwardly prove that this 
condition corresponds to the following identity
\begin{align}\label{eq:no-signalling}
	\begin{aligned}
	\Qcircuit @C=1em @R=1em
	{	&\ustick{\rA}&\multigate{1}{\tU}&\ustick{\rA'}\qw&\measureD{I}\\
		&\ustick{\rB}&\ghost{\tU}&\ustick{\rB'}\qw}
	\end{aligned}
	& =
		\begin{aligned}
	\Qcircuit @C=1em @R=1em
	{	&\ustick{\rA}&\measureD{I}\\
		&\ustick{\rB}&\gate{\tC}&\ustick{\rB'}\qw
	}
	\end{aligned}\quad,
\end{align}
for some channel $\tC\in\QC{\rB}{\rB'}$, where the trivial POVM $I$ on system $\rA$ 
(or $\rA'$) in the diagram represents the partial trace operator $\Tr_{\rA}$ 
(or $\Tr_{\rA'}$) that describes discarding $\rA$ (or $\rA'$). On this basis, given a 
channel $\tU$, we quantify its signalling from $\rA$ to $\rB'$ via the function
\begin{align}
\label{eq:qsig}
S(\tU)\coloneqq\inf_{\tC\in\QC{\rB}{\rB'}}\norma{(\Tr_{\rA'}\otimes \tI_{\rB'})\tU-\Tr_{\rA}\otimes\;\tC}_{\diamond},
\end{align}
where $\tI_{\rB'}$ denotes the
identity channel on system $\rB'$,
$\normad{\tX}\coloneqq\sup_{\rE}\sup_{\rho\in\st(\rE\rA)}\norma{(\tI_{\rE}\otimes\tX)(\rho)}_1$
is the \emph{diamond norm} of the hermitian-preserving map $\tX$ in the real span of $\QC{\rA}{\rA'}$, and $\|\cdot\|_1$ denotes the trace-norm on the space of operators on the Hilbert space $\rE\otimes\rA'$, i.e.~$\|X\|_1\coloneqq\Tr[(X^\dag X)^{1/2}]$.

The signalling condition thus boils down to the possibility of using
$\tU$ to send a message from Alice to Bob, but in a general theory of information 
processing this does not exhaust the ways in which an intervention on system $\rA$ can 
causally affect the system $\rB'$. Indeed a local operation involving only system
$\rA$ before the reversible transformation $\tU$ can influence the output correlations 
between Alice and Bob. This possibility has been extensively explored in 
Refs.~\cite{Perinotti2020cellularautomatain,Perinotti2021causalinfluencein} and 
encompassed in the notion of \emph{causal influence} of system $\rA$ on system
$\rB'$. The definition (by negation) of causal influence is the following. Given the 
unitary $\tU\in\QC{\rA\rB}{\rA'\rB'}$, system $\rA$ has \emph{no causal influence} on 
$\rB'$ if for every $\tA\in\QCs{\rA}$ one has
\begin{align}
  &\begin{aligned}
    \Qcircuit @C=1em @R=1em
    {&\ustick{\rA'}&\multigate{1}{\tU^{-1}}&\ustick{\rA}\qw&\gate{\tA}&\ustick{\rA}\qw&\multigate{1}{\tU}&\ustick{\rA'}\qw\\
      &\ustick{\rB'}&\ghost{\tU^{-1}}&\qw&\ustick{\rB}\qw&\qw&\ghost{\tU}&\ustick{\rB'}\qw}
\end{aligned}
&=
\begin{aligned}	
\Qcircuit @C=1em @R=1em
	{	&\ustick{\rA'}&\gate{\tA'}&\ustick{\rA'}\qw\\
		&&\ustick{\rB'}\qw&\qw
	}
	\end{aligned}
\label{eq:noci}
\end{align}
for a suitable local operation $\tA'\in\QCs{\rA'}$. The above condition
has been proved~\cite{Perinotti2021causalinfluencein} to be strictly stronger than no-signalling for a
general information theory. Indeed on one hand it prevents Alice to
signal to Bob, but it also ensures that the evolution $\tU$ cannot
``propagate'' the effect of any local operation of Alice (on system $\rA$) to
alter the correlations with the output system of Bob's created by $\tU$. Remarkably, in Ref.~\cite{Perinotti2021causalinfluencein} it was also proved that in 
quantum theory no-causal influence coincides with no-signalling, while in classical
information theory there exist examples of channels that cannot be used for transmitting 
signals to a given subsystem but still can be used to influence its correlations. In other 
words, there exist no-signalling gates that have causal influence. As proved in 
Ref.~\cite{Perinotti2021causalinfluencein}, to verify if a channel has
causal influence from $\rA$ to $\rB'$ it is not necessary to check the factorization on 
the rhs of Eq.~\eqref{eq:noci} for every local map $\tA$, but it is sufficient to do it 
on a single probe corresponding to the swap operator between two copies of
Alice's input system $\rA$: in formula, $\tU$ has no causal influence from $\rA$
to $\rB'$ if and only if
\begin{equation}
\label{eq:noci-equivalent}
\begin{aligned}
&{\tT}(\tU)=\tT'\otimes\tI_{\rB'},\\
&{\tT}(\tU)\coloneqq(\tI_\rA\otimes\tU)(\tS\otimes\tI_{\rB})(\tI_{\rA}\otimes\tU^{-1}),
\end{aligned}
\end{equation}
where $\tS\in\QCs{\rA\rA}$ is the swap channel given by
$\tS(\rho)\coloneqq S\rho S$, with
$S\ket{\psi}\otimes\ket{\phi}=\ket{\phi}\otimes\ket{\psi}$ for any
pair $\ket{\phi},\ket{\psi}\in\rA$, and $\tT'$ is a suitable channel in
$\QCs{\rA\rA'}$. We exploit this criterion to define a
quantifier for the causal influence from $\rA$ to $\rB'$ via the
following function
\begin{align}
\label{eq:qcau}
C(\tU)\coloneqq\inf_{\tT'\in\QCs{\rA\rA'}}\norma{{\tT}(\tU)-\tT'\otimes\tI_{\rB'}}_{\diamond}.
\end{align}

We are now in position to compare the two quantities
$S(\tU)$ and $C(\tU)$. 
As we mentioned earlier, a non trivial fact about quantum theory is the equivalence 
between no-signalling and no-causal influence, that can now be expressed as
\begin{equation}
\label{eq:no-int}
  S(\tU)=0\Leftrightarrow C(\tU)=0.
\end{equation}
It is interesting to 
observe a striking consequence of
Eq.~\eqref{eq:no-int}. We know that causal influence includes
signalling as a special case, keeping track also of the correlations
that the channel $\tU$ can generate between Bob's and Alice's
systems at its outcome. On one side it is possible to have signalling
without inducing any correlations, an elementary example being
$\tU\in\QUs{\rA\rB}$ with $\rA\equiv\rB$ and $\tU=\tS$ coinciding with
the swap gate: while signalling from Alice to Bob (and viceversa) is obvious,
since $\tU$ exchanges their systems, if $\rA$ and $\rB$ are
uncorrelated at the input they will remain uncorrelated after the
swap. On the other hand, a channel $\tU$ cannot generate correlations between
Alice and Bob without allowing also for signalling: it is impossible to have 
$C(\tU)\geq 0$ and $S(\tU)=0$ simultaneously.

The first question answer in this Letter is whether the above
equivalence~\eqref{eq:no-int} between no-signalling and no-causal
influence is robust to perturbations of the ideal case where the channel
does not mediate causal relations.  State of the art knowledge on
this subject is null as, in principle, the relative magnitude of the
two quantities may arbitrarily fluctuate as one departs from the condition 
expressed in Eq.~\eqref{eq:no-int}.

This is indeed not the case, as our result is the bound
\begin{align}\label{eq:int-1}
S(\tU)\leq C(\tU)\leq 2\sqrt2 S(\tU)^{\frac{1}{2}}.
\end{align}
These inequalities, proved in the following, establish the robustness of the equivalence
between signalling and causal influence, that can be summarised in the sentence
``little signalling is equivalent to little causal influence''.

The main tool in order to prove Eq.~\eqref{eq:int-1} is a lemma grounding on the continuity of Stinespring
dilations~\cite{c8814692-9976-379a-9ad9-042fab94d853} for quantum channels, 
that we restate in the following, in a slightly different form with respect to the 
original one, for the convenience of the reader. For any quantum channel $\tC\in\QC{\rA}{\rB}$ the
Stinespring theorem implies the existence of a system $\rE$ and an
isometry $\tV\in\QC{\rA}{\rB\rE}$ such that
$\tC=(\Tr_\rE\otimes\tI_{\rB})\circ\tV$. The Stinespring dilation is
charaterized by continuity~\cite{4475375}, namely one can find
dilations of two channels that are close, if and only if the channels
themselves are close.  More precisely given two channels $\tC_1,\tC_2$
and $\tV_1$, $\tV_2$ two of their Stinespring dilations with the same
ancillary system $\rE$, one has~\footnote{The non-trivial part of the
  Stinespring continuity theorem---whose proof can be found in
  Ref.~\cite{4475375}---consists in proving that
\[
  \inf_{U}\norma{(I\otimes U)V_1 - V_2}^2_{\infty} \leq \normad{\tC_1
    - \tC_2}
\]
where $V_1$ and $V_2$ are the Kraus operators of the isometric  dilations of $\tC_1$ and
$\tC_2$ respectively: $ \tC_i(\rho)=\Tr_{E}(V_i\rho V_i^{\dagger})$, $\forall
  \rho\in\st(\rA)$. If $U$ denotes the unitary Kraus operator
of a reversible channel $\tU$, one has
\begin{align*}
  \normad{(\tI_{\rB}\otimes\tU)\tV_1-\tV_2}\leq 2
  \norma{(I_{\rB}\otimes U)V_1-V_2}_{\infty}.
\end{align*}
Indeed, consider a state $\rho\in\st(\rA\rF)$ with
$\rF$  an arbitrary ancillary system, and define $\tV_1^U\coloneqq(\tI_{\rB}\otimes\tU)\tV_1$ with
$V_1^U\coloneqq(I_{\rB}\otimes U)V_1$. Using the fact that
$\norma{ABC}_1\leq \norma{A}_\infty\norma{B}_{1}\norma{C}_\infty$,
$\norma{V}_1=\norma{V^{\dagger}}_1=1$, and the triangle inequality, one can easily check that 
\[
\norma{\tI_{\rF}\otimes[\tV_1^U-\tV_2](\rho)}_1
  \leq  2\norma{V_1^U-V_2}_\infty.
\]
Since this holds for any $\rF$ and any $\rho$, the thesis
follows. As to the second inequality, it is simply due to monotonicity of the diamond norm under partial trace.}
\begin{equation}
\label{eq:Stinespring_cont}
\begin{aligned}
  \inf_{\tU\in\QUs{E}}\normad{(\tU\otimes\tI)\tV_1 - \tV_2}^2&\leq 4
  \norma{\tC_1-\tC_2}_{\diamond} \\
  &\leq 4
  \inf_{\tU\in\QUs{\rE}}\norma{(\tU\otimes\tI)\tV_1 - \tV_2}_{\diamond}.
\end{aligned}
\end{equation}



Thanks to the continuity of Stinespring dilations, we prove the
following bound for a channel
$\tC\in\QC{\rA\rB}{{\rA}^\prime{\rB}}$:
\begin{equation}\label{eq:lemma-1}
\begin{aligned}
  \inf_{\tD\in\QC{\rA}{\rA'}}\normad{\tC-\tD\otimes\tI_{\rB}}^2
  \leq
  4\norma{(\Tr_{\rA'}\otimes\tI_{\rB})\tC-\Tr_{\rA}\otimes\tI_{\rB}}_{\diamond}.
\end{aligned}
\end{equation}

To show inequality~\eqref{eq:lemma-1} consider arbitrary Stinespring
dilations of $\tC$, say $\tV\in\QC{\rA\rB}{\rA'\rB\rE}$ and of $\tD$, say ${\tW}\in\QC{\rA}{\rA'\rE}$,
with auxiliary system $\rE$, and notice that this is also a
Stinespring dilation of $(\Tr_{\rA'}\otimes\tI_{\rB})\tC$ with
auxiliary system $\rA'\rE$. 
Notice also that ${\tW}\otimes\tI_{\rB}$ it
is a Stinespring dilation of $\Tr_A\otimes\tI_\rB$ with auxiliary system
$\rA'\rE$. Therefore, by the first bound in Eq.~\ref{eq:Stinespring_cont}, we have
\begin{align*}
 \inf_{{\tU}\in\QUs{\rA'E}}\normad{({\tU}\otimes\tI_{\rC})\tV-{\tW}\otimes\tI_\rB}^2 \\
= \inf_{{\tU}\in\QUs{\rA'E}}\normad{\tV-{\tU}^{-1}{\tW}\otimes\tI_\rB}^2\\
\leq \  4 \norma{(\Tr_{\rA'}\otimes\tI_{\rB})\tC - \Tr_{\rA}\otimes\tI_\rB}_{\diamond}.
\end{align*}
Defining $\QC{\rA}{\rA'}\ni\tD\coloneqq(\tI_{\rA'}\otimes\Tr_{\rE}){\tU}^{-1}{\tW}$
and using the monotonicity of the diamond norm with respect to the
partial trace, we finally get Eq.~\eqref{eq:lemma-1}.

We can now prove Eq.~\eqref{eq:int-1}. Let us start by proving that
$C(\tU)\leq 2\sqrt2(S(\tU))^{1/2}$. The inequality in
\eqref{eq:lemma-1}, with $\tT(\tU)$ and $\tT'$ playing the role of
$\tC$ and $\tD$, respectively, implies that
\begin{align*}
C^2(\tU)
\leq &  4  \norma{(\Tr_{\rA\rA'}\otimes \tI_{\rB'}){\tT}(\tU) - \Tr_{\rA\rA'}\otimes \tI_{\rB'}}_{\diamond}\\
=& 4  \normad{(\Tr_{\rA\rA'}\otimes \tI_{\rB'})[(\tI_{\rA}\otimes\tU)(\tS\otimes\tI_{\rB})-\tI_{\rA}\otimes\tU]},
\end{align*}
where the equality follows by substituting the explicit expression for
$\tT(\tU)$ in Eq.~\eqref{eq:noci-equivalent} and using the invariance
of the norm with respect to composition with unitary channels. Within
the norm we can add and subtract the term $\Tr_{\rA\rA'}\otimes\tD$
for $\tD\in\QC{B}{B'}$ an arbitrary channel, and use the triangular
inequality together with the properties of the swap transformation to
get
\begin{align*}
  &C^2(\tU)\leq 
8 \ \normad{(\Tr_{\rA'}\otimes \tI_{\rB'})\tU-\Tr_{\rA}\otimes\tD}.
\end{align*}
Finally, since the above inequality holds for
every $\tD$, it also holds for the infimum over $\tD\in\QC{B}{B'}$, which
concludes the proof.

We now show the other bound $S(\tU)\leq  C(\tU)$. For an arbitrary $\tT'\in\QUs{\rA\rA'}$ one has
\begin{align*}
\begin{aligned}
\normad{{\tT}(\tU)&-\tT'\otimes\tI_{\rB'}}
\\
\geq &\ \normad{(\Tr_{\rA\rA'}\otimes \tI_{\rB'})[(\tI_{\rA}\otimes\tU)(\tS\otimes\tI_{\rB}) \\
 &-(\tT'\otimes\tI_{\rB'})(\tI_{\rA}\otimes\tU)](\tI_{\rA}\otimes\rho\otimes\tI_{\rB})},
\end{aligned}
\end{align*}
where the inequality follows from the monotonicity of the
norm with respect to the partial trace $\Tr_{\rA\rA'}$, and with respect to preparation of 
a fixed state $\rho$ of system $\rA$, along with the explicit 
form of ${\tT}(\tU)$ given in Eq.~\eqref{eq:noci-equivalent} and invariance of
the norm under composition with unitary channels. Observing that
$(\Tr_{\rA\rA'}\otimes
\tI_{\rB'})(\tI_{\rA}\otimes\tU)(\tS\otimes\tI_{\rB})(\tI_{\rA}\otimes\rho\otimes\tI_{\rB})=(\Tr_{\rA'}\otimes\tI_{\rB'})\tU$,
and defining
\begin{align*}
\tD\coloneqq(\Tr_{\rA'}\otimes \tI_{\rB'})\tU(\rho\otimes\tI_{\rB}),
\end{align*}
we conclude that for every $\tT'$ there exists $\tD$ such that
\begin{align*}
 \normad{{\tT}(\tU)-\tT'\otimes\tI_{\rB'}}\geq  \normad{(\Tr_{\rA'}\otimes\tI_{\rB'})\tU - \Tr_{\rA}\otimes \tD },
\end{align*}
thus proving the desired relation.

The core message of the bounds between causal influence and signalling
is that if one of them is small the other is also small. Due to
$S(\tU)\leq C(\tU)$, if a reversible channel $\tU$ allows for a little
bit of causal influence, say no more than $\varepsilon$, then also the
amount of signalling that is allowed is bounded by
$\varepsilon$. Conversely, from $C(\tU)\leq 2\sqrt2 S(\tU)^{1/2}$, if
$\tU$ allows for a small, say $\varepsilon$, signalling, then it
cannot exhibit causal influence bigger than $2\sqrt{2\varepsilon}$. 

Notice however that, due to the singularity of the derivative of $x^{1/2}$ in $x=0$, 
in a neighbourhood of $S(\tU)=0$, one can have a large increase in causal influence 
with a negligible increase in signalling. This very observation can be seen as 
spotlighting the remnant of the non-equivalence of the two notions that we remarked 
in the classical case.

The second main result of this Letter is indeed the proof that causal
influence and signalling are not equal. The mismatch between the
two quantities is definitely established by their analytical
computation, provided in the following, for the quantum
$\mathrm{Cnot}$ unitary channel:
\begin{align}\label{eq:cnot}
C(\mathrm{Cnot})=2>1\geq S(\mathrm{Cnot}).
\end{align}
This states that there exist interactions where Alice's local
operations have effects on Bob's system that ``exceed'' those on Bob's local states. Such effects are not to be sought in communication
capacity but in the perturbation of Bob's system correlations.

We prove Eq.~\eqref{eq:cnot} computing the norms in
Eqs.~\eqref{eq:qsig} and~\eqref{eq:qcau} for $\tC=\mathrm{Cnot}$. In
the present case it is
$\rA \cong\rB \cong\rA' \cong\rB' \cong\mathbb{C}^2$, but we keep general
labels for input and output systems for ease of comparison
with the above definitions. To prove that $C(\mathrm{Cnot})=2$ we
show that for any choice of channel $\tT'$, one can always find a
state $\rho\in\st{(\rE\rA\rA'\rB')}$ such that
\begin{align*}
 \| (\tI_E\otimes \tT (\mathrm{Cnot}) - \tI_E\otimes \tT'
  \otimes \tI_{B'})(\rho)\|_1=2.
\end{align*}
Let $\rho =\proj{\psi}$ with
$\ket{\psi}=\ket{+-++}$ and
$\ket{\pm}:=(\ket{0}\pm\ket{1})/\sqrt{2}$. Remembering the
 definition of $\mathrm{Cnot}$
   ($\mathrm{Cnot}(\proj{ab})=\proj{(a\oplus b)b}$, with
   $a,b\in\{0,1\}$ and where the target
   qubit is the first qubit), one can easily verify that
$(\tI_E\otimes \tT(\mathrm{Cnot}))(\rho)=\proj{\psi'}$ with
$\ket{\psi'}=\ket{++--}$, whereas
$(\tI_E\otimes \tT' \otimes \tI_{B'})(\rho)=\proj{+} \otimes
\tT'(\proj{-+})\otimes \proj{+}$. The two output states of the last qubit ($\rB'$) are
perfectly distinguishable as one can see by direct inspection. We remind indeed that for 
any pair of states $\sigma$ and $\nu$,
it is
$\left\|\sigma - \nu \right\|_1 = 2 \max_P \Tr[P(\sigma - \eta)]$
where the maximum is taken over all projectors. It is then sufficient to choose the
projector $P=I_{EAA'}\otimes \proj{+}$, and conclude that
$\left\|\proj{\psi'} - \proj{+} \otimes
  \tT'(\proj{-+})\otimes \proj{+} \right\|_1 = 2$, thus
proving the thesis.

We finally show that $S(\mathrm{Cnot})\leq1$. To this end it suffices
to exhibit a channel $\tM\in\QC{\rB}{\rB'}$ such that
\begin{align*}
S_\tM(\mathrm{Cnot}):=\norma{(\Tr_{\rA'}\otimes \tI_{\rB'})\mathrm{Cnot}-\Tr_{\rA}\otimes\;\tM}_{\diamond}=1.
  \end{align*}
  Indeed according to defintion~\eqref{eq:qsig} one has
  $S(\mathrm{Cnot})\leq S_\tM(\mathrm{Cnot})$ for every
  $\tM\in\QC{\rB}{\rB'}$.  Let then $\tM$ be the
  channel that performs a non-selective measurement in the computational basis,
  namely   $\tM : \rho \mapsto P_0 \rho P_0 + P_1 \rho P_1$,
 with
  projectors $P_1:=\proj{0}$ and $P_1:=\proj{1}$, and show that
  $S_{\tM}(\mathrm{Cnot})=1$. By definition of diamond norm one has
\begin{align*}
 & S_{\tM}(\mathrm{Cnot})\\
  &=    \sup_{\rho}\|(\tI_\rE \otimes(\Tr_{\rA'}\otimes \tI_{\rB'}) \mathrm{Cnot}  - \tI_\rE \otimes \Tr_{\rA}\otimes \tM )(\rho) \|_1 \\
  &=    \sup_{\rho}\|(\tI_\rE \otimes\Tr_{\rA}\otimes \tI_{\rB} - \tI_\rE \otimes (\Tr_{\rA}\otimes \tM)  \mathrm{Cnot} )(\rho) \|_1,
\end{align*}
 where we replaced $\rho\in\st(\rE\rA\rB)$ with
 $(\tI_\rE\otimes \mathrm{Cnot})(\rho)$, which also ranges over
 $\st(\rE\rA\rB)$, and used the fact that $\mathrm{Cnot}$ is its own
 inverse.  A simple computation shows that
 $(\Tr_{\rA}\otimes \tM)\mathrm{Cnot} = \Tr_{\rA}\otimes \tM$, which
 can be checked on pure states using the defintion of $\mathrm{Cnot}$.
 Therefore one has the following simpler expression for
 $S_{\tM}(\mathrm{Cnot})$:
 \begin{align*}
  S_{\tM}(\mathrm{Cnot})= \sup_{\rho \in\st(\rE\rB)} \| \rho - (\tI_\rE \otimes \tM )(\rho) \|_1.
 \end{align*}
 By convexity of the trace norm it is sufficient to take the supremum
 over pure states, and we can now assume that $\rE \cong \rB$. Let
 then $\rho\in\st(\rE\rB)$ be a pure state, which we write as
 $\rho =\proj{\psi}$ with
 $\ket{\psi}=\alpha
 \ket{00}+\beta\ket{01}+\gamma\ket{10}+\delta\ket{11}$ and $|\alpha|^2+|\beta|^2+|\gamma|^2+|\delta|^2=1$. We observe that
 $\left\| \rho - \bigl(\mathcal{I}_E \otimes \mathcal{M} \bigr)(\rho)
 \right\|_1$ can be expressed as the sum of the absolute values of the
 eigenvalues of
 $\rho - \bigl(\mathcal{I}_E \otimes \mathcal{M} \bigr)(\rho)$. The
 latter has nonzero eigenvalues $\pm\sqrt{p(1-p)}$ with
 $p:=|\alpha|^2+|\gamma|^2$ and $1-p:=|\beta|^2+|\delta|^2$,
 corresponding to the probability that the measurement $\tM$ outcome
 is 0 and 1, respectively. Accordingly we finally get
 \begin{align*}
 \left\| \rho - \bigl(\mathcal{I}_E \otimes \mathcal{M} \bigr)(\rho) \right\|_1 = 2\sqrt{p(1-p)},
\end{align*}
that is maximal for $p=\frac{1}{2}$. In this case
$\| \rho - (\tI_E \otimes \tM )(\rho)\|_1 =1$, which concludes the proof.

\emph{Conclusion and discussion.}---In this Letter, we have shown how
the full amount of causal relations---say the causal
influence---activated by a quantum unitary channel is strictly larger
than the fraction of causal relations represented by its
signalling. The $\mathrm{Cnot}$ gate is an example in which a lower
bound on the gap
between the two quantities can be analytically computed and is proved
to be non-vanishing. We proved that causal influence scales as a
function of signalling. While no interaction---i.e.~free evolution---is equivalent
to the absence of both causal influence and communication, a unitary
coupling between quantum systems generates a little causal influence
if and only if it allows for a little signalling.

The strength of causal influence of a unitary channel has been
quantified in strictly operational terms within the channels
discrimination framework. A similar approach has
been recently adopted in Ref.~\cite{Cotler:2019aa}, where the authors
study the emergence of causal relations between spatial regions of
spacetime as induced by an Hamiltoninan evolution. However, the notion
of causal influence defined in Ref.~\cite{Cotler:2019aa} is relative
to a fixed initial state of the quantum network while the present
approach is state-independent. 

As the information exchanged between different parties is typically
quantified via entropy-based measures, a future perspective of the present results is
to generalise them to the analysis the behaviour of some significant entropic function.
A natural candidate in this direction is the \emph{entropy exchange}
between the input held by Alice and Bob's output. Such an entropic
characterization of causal influence will be of interest for
cryptographic applications and for investigating the causal structure
of quantum many-body systems. 



\emph{Acknowledgements.} {P. P. acknowledges financial support from PNRR MUR
  project PE0000023-NQSTI. A. T. acknowledges the financial support of Elvia
  and Federico Faggin Foundation (Silicon Valley Community Foundation
  Project ID\#2020-214365).}

\bibliography{bibliography}
\bibliographystyle{apsrev4-2.bst}

\end{document}